\documentclass[aps,pra]{revtex4}
\usepackage{amsmath}
\usepackage{amsfonts}
\usepackage{amssymb}
\usepackage{bm}
\usepackage{graphicx}
\begin{document}

\title{Extended Korteweg-de Vries equation for long gravity waves in incompressible fluid without  strong limitation to surface deviation}
\affiliation{Centre for Engineering Quantum Systems, School of Mathematics and Physics,
The University of Queensland, Brisbane, Queensland 4072, Australia}
\author{ Vladimir I. Kruglov}
\affiliation{Centre for Engineering Quantum Systems, School of Mathematics and Physics, The University of Queensland, Brisbane, Queensland 4072, Australia}

\begin{abstract}
	
We have derived the extended Korteweg-de Vries equation describing the long gravity waves without limitation to surface deviation. The only restriction to the surface deviation is connected with the stability condition for appropriate solutions. The derivation of extended  KdV equation is based on the Euler equations for inviscid irrotational and incompressible fluid. It is shown that the extended KdV equation reduces to standard KdV equation for small amplitude of the waves. We have also generalized the extended KdV equation for describing the decaying effect of the waves. Quasi-periodic and solitary wave solutions for extended KdV equation with decaying effect are found as well.  We also demonstrate that the fundamental approach based on the inverse scattering method is applicable for solving the extended KdV equation in the case when decaying effect is negligibly small. Such case always occur for restricted propagation distances of the waves.   
	
\end{abstract}

%\pacs{}
\maketitle

\section{Introduction}

The Korteweg-de Vries equation (KdV) describes the 
shallow water waves with small but finite amplitude \cite{1,2, 3,4}. It is one of the most successful physical equation consisting the simplest possible terms representing the interplay of dispersion and nonlinearity.  
The KdV equation also describes pressure waves in a bubble-liquid mixture \cite{5}; acoustic waves and heat pulses in anharmonic crystals \cite{6,7,8};
magnetic-sonic waves in magnetic plasma \cite{9,10,11,12}; electron plasma waves in a cylindrical plasma \cite{13,14}; and ion acoustic waves \cite{15,16,17,18}. The derivation of KdV equation for enough general class of equations is given in \cite{19,20}. Zabusky and Galvin 
\cite{21} have shown that KdV equation leads to very accurate description for weakly decaying waves propagating in shallow water. Numerous results for the KdV equation have been obtained in recent years. The important methods and results are given by Gardner, Green, Kruskal, Lax, Miura, Hirota, and others in Refs.\cite{22,23,24,25,26,27}. Many other impotent results for the Korteweg-de Vries equation have also presented for an example in Refs. \cite{28,29,30,31}.
The KdV equation is tested experimentally as a model for 
moderate amplitude waves propagating in one direction in relatively shallow water of uniform depth. For a wide range of initial data, comparisons are made between the asymptotic wave forms observed and those predicted by the theory in terms of the number of solitons that evolve, the amplitude of the leading soliton, the asymptotic shape of the wave and other qualitative features \cite{32}. Computations made in this work by Hammack and Segur suggest that the KdV equation predicts the amplitude of the leading soliton to within the 
expected error due to viscosity (12$\%$) when the non-decayed amplitudes are less than about a quarter of the water depth. The agreement to within about 20$\%$ 
is observed over the entire range of experiments examined, including those with initial data for which the non-decayed amplitudes of the leading soliton exceed half the fluid depth.

The purpose of present paper is derivation of the extended KdV equation for gravity waves in compressible fluid without  restriction to amplitude of the waves. Moreover, the derived extended KdV equation is  generalized to describe the decaying effect for the gravity waves. This extended KdV equation is found for the long waves or shallow fluid. The long wavelength condition is similar to the appropriate condition used in derivation of the standard Korteweg-de Vries equation \cite{1,2, 3,4}.
We emphasize that in our derivation of the extended KdV equation it is not assumed the small wave amplitude condition  $\vert\eta\vert/h_{0}\ll 1$ where $\eta(x,t)$  is the surface deviation of the waves under an equilibrium level $h_{0}$.  The only limitation for wave amplitude in the extended KdV equation is connected with the stability condition for the gravity waves. Note that the derived extended KdV equation without decaying effect reduces to the KdV equation when the additional condition $\vert\eta\vert/h_{0}\ll 1$ is satisfied.
It is shown that the term describing decaying effect in the extended KdV equation depends on two parameters as the kinematic viscosity $\nu$ (momentum diffusivity) and the capillary length $\lambda_{c}$. The explicit form for the decaying term is derived using the dimensionless analysis with the critical parameters $\nu$ and $\lambda_{c}$. Using the perturbation method we have found the set of decaying quasi-periodic and solitary wave solutions for extended KdV equation. We also demonstrate that the fundamental approaches based on the inverse scattering method are applicable for solving the extended KdV equation in the cases when decaying effect is negligibly small. Such cases always occur for restricted propagation distances of the gravity waves.    

The results in this paper are presented as follows. Sec. II presents the derivation of extended KdV equation. In Sec. III, we consider the propagation of traveling gravity waves in shallow water with an arbitrary amplitude. In Sec. IV,
we generalize the extended KdV equation for describing the decaying effect of gravity waves. The decaying traveling wave solutions for extended KdV equation are obtained in Sec. V. In Sec. VI, we present the discussion of obtained decaying wave solutions. Finally, we summarize the results in Sec. VII.

\section{Extended KdV equation for long gravity waves}

The waves in shallow water of uniform depth is described by the Euler equations for inviscid and incompressible fluid together with  conservation equation:
\begin{equation}
\partial_{t}u+u\partial_{x}u+w\partial_{z}u=-\frac{1}{\rho}\partial_{x}P,\label{1}
\end{equation}%
\begin{equation}
\partial_{t}w+u\partial_{x}w+w\partial_{z}w=-\frac{1}{\rho}\partial_{z}P-g,
\label{2}
\end{equation}%
\begin{equation}
\partial_{x}u+\partial_{z}w=0,
\label{3}
\end{equation}
where $\mathbf{v}=(u, 0, w)$ is the velocity, $P$ is pressure, $g$ is the acceleration by gravity, and we assume that $\rho=const$. The condition $\nabla\times\mathbf{v}=0$ can be used to introduce the potential of velocity as $\mathbf{v}=\nabla\phi$. Thus we have $u=\partial_{x}\phi$ and $w=\partial_{z}\phi$, and the equation for potential $\phi$ follows from  Eq. (\ref{3}) as $\partial_{x}^{2}\phi+\partial_{z}^{2}\phi=0$.
However, in this paper we don't use the approach based on potential $\phi$ because this function depends on three variables as $x$, $z$ and $t$.
 
We note that the water depth $h$ for waves propagating to x-direction depends on the time $t$ and longitudinal coordinate $x$. Thus, the water depth for the waves is $h(x,t)=h_{0}+\eta(x,t)$ where $\eta(x,t)$ is the surface deviation under the equilibrium level $h_{0}$.
We consider below the propagation of long waves which means that the following condition is satisfied: $\epsilon^{2}\ll 1$ where $\epsilon=h_{0}/l$  and $l$ is the characteristic length of the wave. 
It is shown in the Appendix A (sec.1) that the full pressure $P$ can be presented as the sum of static $P_{g}$ and dynamic $P_{d}$ gravitational pressures respectively. Moreover, the static pressure is given by equation as $\rho^{-1}\partial_{z}P_{g}=-g$. 
Thus, the full pressure $P$ and the static pressure $P_{g}$ are given by
\begin{equation}
P=P_{g}+P_{d},~~~~P_{g}=P_{0}+\rho g[h(x,t)-z],
\label{4}
\end{equation}%
where $z$ is the vertical coordinate and $P_{0}$ is the pressure at $z=h$.  We can present the term  $\rho^{-1}\partial_{x}P$ by Eq. (\ref{4}) as
\begin{equation}
\frac{1}{\rho}\partial_{x}P=g\partial_{x}\eta+\frac{1}{\rho}\partial_{x}P_{d},
\label{5}
\end{equation}
where the dynamic pressure $P_{d}(x,t)$ depends on variables $x$ and $t$. One can also use the standard assumption that the velocity $u=u(x,t)$ depends on variables $x$ and $t$ only. This is correct when the initial velocity $u(x,0)$ does not depend on variable $z$. In this case Eqs. (\ref{1}) and (\ref{5}) lead to the equation,
\begin{equation}
\partial_{t}u+u\partial_{x}u+g\partial_{x}\eta+\mathcal{D}=0,
\label{6}
\end{equation}%
where $\mathcal{D}(x,t)\equiv\rho^{-1}\partial_{x}P_{d}(x,t)$. It is shown below that the term $\mathcal{D}$ is necessary in Eq. (\ref{6}) for correct description of the dispersion relation in the first order to small parameter $\epsilon^{2}$. This explains the insertion of dynamic gravitational pressure $P_{d}$ in Eqs. (\ref{4}) and (\ref{5}).   

The conservation equation (\ref{3}) for gravity waves reduces to standard form as
\begin{equation}
\partial_{t}h+\partial_{x}(uh)=0.
\label{7}
\end{equation}%
The derivation of this conservation equation is presented in Appendix A (sec.2). Thus, the Euler equations for long gravity waves lead to the system of Eqs. (\ref{6}) and (\ref{7}). The explicit form for dynamical pressure $P_{d}$ and the function $\mathcal{D}(x,t)$ is derived below using special transformation and dispersion relation for waves on water surface.  

We have found the following transformation which is important
for derivation of the extended KdV equation:
\begin{equation}
u(x,t)=2\sqrt{gh_{0}+g\eta(x,t)-r(x,t)}-2\sqrt{gh_{0}},
\label{8}
\end{equation}%
where $r(x,t)$ is some new function. This transformation means that the velocity $u(x,t)$ depends on two independent functions as $\eta(x,t)$ and $r(x,t)$.
We emphasize that Eq. (\ref{8}) is not a Riemann invariant for the system of Eqs. (\ref{6}) and (\ref{7}). 
The Riemann invariant of Eqs. (\ref{6}) and (\ref{7}) for condition $\mathcal{D}(x,t)\equiv 0$ is presented in Appendix A (sec.3). The transformation given in Eq. (\ref{8}) can also be written as
\begin{equation}
\eta(x,t)=\frac{1}{g}r(x,t)+\frac{c_{0}}{g}u(x,t)
+\frac{1}{4g}u^{2}(x,t),
\label{9}
\end{equation}
where $c_{0}=\sqrt{gh_{0}}$ is the characteristic velocity. This
characteristic velocity is connected with dispersion equation for the waves on water surface.
Applying the transformation in Eq. (\ref{9}) to system of Eqs. (\ref{6}) and (\ref{7}) we have found the following system of equations,
\begin{equation}
\partial_{t}u+c_{0}\partial_{x}u
+\frac{3}{2}u\partial_{x}u+\partial_{x}r+\mathcal{D}=0,
\label{10}
\end{equation}
\begin{equation}
\partial_{t}r+\partial_{x}(ur)=\left(c_{0}
+\frac{1}{2}u\right)(\partial_{x}r+\mathcal{D}).	
\label{11}
\end{equation}%

The dispersion relation for waves on water surface \cite{33} is
\begin{equation}
\omega^{2}=\left(1+\frac{\gamma\kappa^{2}}{\rho g}\right) g\kappa\tanh(\kappa h_{0}).
\label{12}
\end{equation}%
We have in the case $\kappa^{2}h_{0}^{2}\ll 1$ the following decomposition, 
\begin{equation}
\omega=c_{0}\kappa-\left(\frac{h_{0}^{2}}{6}-\frac{\gamma}{2\rho g}  \right)c_{0}\kappa^{3}+...~,
\label{13}
\end{equation}%
where $\kappa$ and $\gamma$ are the wave number and surface tension respectively. The first two terms in this dispersion equation are found in the first order to small parameter $\kappa^{2}h_{0}^{2}$. The wave number can be written as $\kappa\approx 1/l$ where $l$ is a characteristic length of the wave propagating to the $x$-direction. Thus, the first two terms
in Eq. (\ref{13}) are given in the first order to small parameter
$\epsilon^{2}$. We require that dispersion equation in Eq. (\ref{13}) (with the first two terms) follows from linearized Eq. (\ref{10}). The only linear differential equation for the function $u(x,t)$ satisfying to this condition has the form,  
\begin{equation}
\partial_{t}u+c_{0}\partial_{x}u+\sigma\partial_{x}^{3}u=0,
\label{14}
\end{equation}
where the parameter $\sigma$ is 
\begin{equation} 
\sigma=\frac{c_{0}h_{0}^{2}}{6}-\frac{c_{0}\gamma}{2\rho g}.	
\label{15}
\end{equation}
This result follows by substitution of the plain wave $u=A\exp[i(\kappa x-\omega t)]$ to Eq. (\ref{14}).

We note that Eq. (\ref{10}) depends on the function $r(x,t)$ which one can consider in the form $r=f(u)$ where the function $f(u)$ is defined by Eqs. (\ref{10}) and (\ref{11}).
The linearized  function $r=f(u)$ has the form $r=\alpha+\nu u$ 
where $\alpha$ and $\nu$ are some unknown coefficients.
The substitution of linearized  function $r=\alpha+\nu u$ to  Eq. (\ref{9}) yields $\alpha=0$ because $u\equiv 0$ in the case when $\eta\equiv 0$. Thus, in the general case linearized fuction $r=f(u)$ has the form $r=\nu u$ which leads to linearized Eq. (\ref{10}) as
\begin{equation}
\partial_{t}u+(c_{0}+\nu)\partial_{x}u+\mathcal{D}=0.
\label{16}
\end{equation}%
We claim that Eqs. (\ref{14}) and (\ref{16}) are equivalent equations which leads to coefficient $\nu=0$ and the last term in the left side of (\ref{16}) is $\mathcal{D}=\sigma\partial_{x}^{3}u$. Hence, we have the function $\mathcal{D}$ and dynamic pressure $P_{d}$ as
\begin{equation}
\mathcal{D}(x,t)=\sigma\partial_{x}^{3}u(x,t),~~~~
P_{d}(x,t)=\rho\sigma\partial_{x}^{2}u(x,t),
\label{17}
\end{equation}
where $\sigma=(\chi/6) c_{0}h_{0}^{2}$ and $\chi=1-3\gamma/\rho gh_{0}^{2}$. We note that the pressure $P_{d}$ is found by relation $\rho^{-1}\partial_{x}P_{d}\equiv\mathcal{D}$ and the full pressure is given by equation as $P=P_{g}+P_{d}$. Thus, we have found the full system of Eqs. (\ref{6}) and (\ref{7}) where the term $\mathcal{D}=\sigma\partial_{x}^{3}u$ is connected with the dynamic pressure $P_{d}$.

We define the dimensionless variables 
$\tau=c_{0}t/l$, $\lambda=x/l$, and dimensionless functions $\tilde{u}$, $\tilde{r}$ and $\tilde{\eta}$ as
\begin{equation}
\tilde{u}(\lambda,\tau)=\frac{1}{c_{0}}u(x,t),~~~~
\tilde{r}(\lambda,\tau)=\frac{1}{gh_{0}}r(x,t),~~~~\tilde{\eta}=\eta/h_{0}.	
\label{18}
\end{equation}%
Hence, the function $\mathcal{D}(x,t)$ given in Eq. (\ref{17}) has the form,
\begin{equation}
\mathcal{D}(x,t)=\frac{\epsilon^{2}c_{0}^{2}\chi}{6l}\partial_{\lambda}^{3}\tilde{u}(\lambda,\tau).
\label{19}
\end{equation}%
This equation can also be written as
\begin{equation}
\mathcal{D}(x,t)=\frac{c_{0}^{2}}{l}\tilde{\mathcal{D}}(\lambda,\tau),~~~~\tilde{\mathcal{D}}(\lambda,\tau)=\frac{\epsilon^{2}\chi}{6}
\partial_{\lambda}^{3}\tilde{u}(\lambda,\tau),
\label{20}
\end{equation}%
where $\tilde{\mathcal{D}}(\lambda,\tau)$ is dimensionless function connected to dynamical pressure $P_{d}$. 
Eqs. (\ref{19}) and (\ref{20}) demonstrate that the term $\mathcal{D}(x,t)$ in Eq. (\ref{6}) has the first order to small parameter $\epsilon^{2}$. 
Eqs. (\ref{10}), (\ref{11}) and (\ref{9}) with dimensionless functions defined in Eqs. (\ref{18}) and (\ref{20}) have the following dimensionless form,
\begin{equation}
\partial_{\tau}\tilde{u}+\partial_{\lambda}\tilde{u}+\partial_{\lambda}\tilde{r}+\frac{3}{2}\tilde{u}\partial_{\lambda}\tilde{u}+\tilde{\mathcal{D}}=0,
\label{21}
\end{equation}%
 \begin{equation}
\partial_{\tau}\tilde{r}+\partial_{\lambda}(\tilde{u}\tilde{r})=\left(1+\frac{1}{2}\tilde{u}\right)\left(\partial_{\lambda}\tilde{r}
+\tilde{\mathcal{D}}\right),	
\label{22}
\end{equation}%
\begin{equation}
\tilde{\eta}=\tilde{r}+\tilde{u}+\frac{1}{4}\tilde{u}^{2}.	
\label{23}
\end{equation}%
We note that the system of Eqs. (\ref{6}) and (\ref{7}) with $\mathcal{D}(x,t)\equiv 0$ has the Riemann invariant given by Eq. (\ref{8}) with $r(x,t)\equiv 0$ [see Appendix A (sec.3)]. Thus, the function $r(x,t)$ arises in Eqs. (\ref{10}) and (\ref{11}) only in the case when $\mathcal{D}(x,t)\neq 0$. Using the perturbation theory to small parameter $\epsilon^{2}$ we can write the dimensionless function $\tilde{r}$ in the form $\tilde{r}(\lambda,\tau)=\tilde{q}_{0}(\lambda,\tau)+\epsilon^{2}\tilde{q}(\lambda,\tau)$ where the function $\tilde{q}(\lambda,\tau)$ is a polynomial to small parameter $\epsilon^{2}$. Eq. (\ref{20}) yields  $\tilde{\mathcal{D}}(\lambda,\tau)=0$ for $\epsilon^{2}=0$, and hence we have $\tilde{r}(\lambda,\tau)=0$ at $\epsilon^{2}=0$. Thus, we have found that $\tilde{q}_{0}(\lambda,\tau)=0$ and the dimensionless function $\tilde{r}$ has the general form as $\tilde{r}(\lambda,\tau)=\epsilon^{2}\tilde{q}(\lambda,\tau)$.
This means that the function $\tilde{r}$ has first order to small parameter $\epsilon^{2}$.
 
We can neglect the terms $\partial_{\lambda}\tilde{r}$ and $\tilde{r}$ in Eq. (\ref{21}) for long wave approximation. However, we don't neglect the term $\tilde{\mathcal{D}}$ in Eq. (\ref{21}) because the dynamical behavior of the gravity waves especially depends on this term connected to dynamic pressure $P_{d}$. It is important that this term leads to correct dispersion equation in the first order to small parameter $\epsilon^{2}$. Moreover, Eq. (\ref{22}) is satisfied for the long wave approximation because the left and right hand sides of this equation are proportional to small parameter $\epsilon^{2}$. We can also neglect the small term $\tilde{r}=\epsilon^{2}\tilde{q}$ in Eq. (\ref{23}).
Hence, the system of Eqs. (\ref{21}), (\ref{22}) and (\ref{23}) reduces to pair equations as
\begin{equation}
\partial_{\tau}\tilde{u}+\partial_{\lambda}\tilde{u}
+\frac{3}{2}\tilde{u}\partial_{\lambda}\tilde{u}+\frac{\epsilon^{2}\chi}{6}\partial_{\lambda}^{3}\tilde{u}=0,~~~~\tilde{\eta}=\tilde{u}+\frac{1}{4}\tilde{u}^{2}.
\label{24}
\end{equation}
The dimensional form for this system of equations can be written as  
\begin{equation}
\partial_{t}u+c_{0}\partial_{x}u
+\frac{3}{2}u\partial_{x}u+\sigma\partial_{x}^{3}u=0,
\label{25}
\end{equation}
\begin{equation}
\eta(x,t)=\frac{c_{0}}{g}u(x,t)
+\frac{1}{4g}u^{2}(x,t),
\label{26}
\end{equation}
where the term $\sigma\partial_{x}^{3}u$ is connected with dynamic pressure $P_{d}$ presented in Eq. (\ref{17}). We note that the derived Eq. (\ref{25}) for velocity $u$ and the Burgers equation \cite{34,35,36} have significantly different form. In particular, the difference in these two cases for terms with higher order derivatives leads to significant various classes of solutions. Eq. (\ref{25}) without the last term $\sigma\partial_{x}^{3}u$ leads to solutions presented in Appendix A (sec.3). The extended KdV Eq. (\ref{25}) is found for long gravity waves without limitation to surface deviation. The only restriction to the surface deviation is connected with the stability condition for appropriate solutions.

The dynamic gravitational pressure can be written by Eqs. (\ref{17}) and (\ref{26}) as
\begin{equation} 
P_{d}=\frac{\chi}{3}\rho gh_{0}^{5/2}\partial_{x}^{2}h^{1/2},  	
\label{27}
\end{equation}
with $h=h_{0}+\eta$.
We also present the system of Eqs. (\ref{25}) and (\ref{26}) in other form introducing new function $\zeta(x,t)$ as
\begin{equation} 
\zeta(x,t)=\frac{c_{0}}{g}u(x,t).
\label{28}
\end{equation}
Thus, the system of Eqs. (\ref{25}) and (\ref{26}) can be written in the following form, 
\begin{equation}
\partial_{t}\zeta+c_{0}\partial_{x}\zeta+\beta\zeta\partial_{x}\zeta+
\sigma\partial_{x}^{3}\zeta=0,
\label{29}
\end{equation}%
\begin{equation}
\eta(x,t)=\zeta(x,t)+\frac{1}{4h_{0}}\zeta^{2}(x,t).
\label{30}
\end{equation}%
The parameters $\beta$ and $\sigma$ in the extended KdV Eq. (\ref{29}) are
\begin{equation} 
\beta=\frac{3c_{0}}{2h_{0}},~~~~\sigma=\frac{\chi c_{0}h_{0}^{2}}{6},~~~~\chi=1-\frac{3\gamma}{\rho gh_{0}^{2}}.
\label{31}
\end{equation}
It is important that the extended KdV equation (\ref{29}) has the same form as the standard KdV equation, however the function $\eta(x,t)$ is given here by Eq. (\ref{30}). Hence, one can apply to extended KdV equation (\ref{29}) all methods developed for solution of the KdV equation. Some exact solutions of the extended KdV equation are given in the Appendix B. 

Thus, we have not used in derived extended KdV Eq. (\ref{29}) the condition $|\eta|/h_{0}\ll 1$ which is a necessary suggestion for the Korteweg-de Vries equation. We have used in our derivation of the system of Eqs. (\ref{29}) and (\ref{30}) the long wave approximation $\epsilon^{2}\ll 1$ only. The experimental observations show that the solitary waves propagating in shallow water are stable when the condition $\eta(x,t)<\eta_{0}$ is satisfied where $\eta_{0}/h_{0}\approx 0.7$  \cite{33}. Our theoretical stability condition for parameter $\delta=\eta_{0}/h_{0}$ is given as $\delta\approx 0.76$ which is close to the experimental observations. We emphasize that the extended KdV equation (\ref{29}) coincides with the KdV equation when the following additional condition $\vert\eta\vert/h_{0}\ll 1$ is satisfied. In this case Eq. (\ref{30}) yields the relation $\zeta(x,t)=\eta(x,t)$ which transforms Eq. (\ref{29}) to KdV equation.

We also note that the pressure $P_{d}$ given by Eq. (\ref{27}) and the function $\mathcal{D}\equiv\rho^{-1}\partial_{x}P_{d}$ used in the extended KdV equation (\ref{29}) are proportional to the functions $\partial_{x}^{2}h^{1/2}$ and $\partial_{x}^{3}h^{1/2}$ respectively. However, the pressure $P_{d}$ connected to the KdV equation is $P_{d}=(\chi/6)\rho gh_{0}^{2}\partial_{x}^{2}\eta$ which follows from Eq. (\ref{17}) with $u=(g/c_{0})\eta$ because in this case $\zeta=\eta$. Hence, the pressure $P_{d}$ and the function $\mathcal{D}$ for KdV equation are proportional to the functions $\partial_{x}^{2}\eta$ and $\partial_{x}^{3}\eta$ respectively. The difference for pressure $P_{d}$ and the function $\mathcal{D}$ in these two cases is crucial for the developed theory of gravity waves in incompressible fluid. In the case when condition $|\eta|/h_{0}\ll 1$ is satisfied the pressure $P_{d}$ and the function $\mathcal{D}$ are the same for extended KdV Eq. (\ref{29}) and the standard KdV equation. This result follows from Eq. (\ref{27}) and decomposition  $h^{1/2}=h_{0}^{1/2}(1+\eta/2h_{0}+...)$ with $|\eta|/h_{0}\ll 1$.

\section{Traveling waves for extended KdV equation}

In this section we consider the propagation of traveling gravity waves in shallow water without standard limitation to
surface deviation. Description of long waves is based here on the extended KdV equation (\ref{29}) and additional relation (\ref{30}) for the function $\eta(x,t)$. Integration of Eq. (\ref{29}) for traveling waves leads to the second order nonlinear differential equation,
\begin{equation} 
2\chi h_{0}^{2}\frac{d^{2}F}{ds^{2}}+9F^{2}
+12\left(1-\frac{v_{0}}{c_{0}}\right)F
+12C_{1}=0,
\label{32}
\end{equation}
where $F(s)=h_{0}^{-1}\zeta(x,t)$, $s=x-v_{0}t$ and $C_{1}$ is integration constant. The second integration yields the first order nonlinear differential equation as
\begin{equation} 
\chi h_{0}^{2}\left(\frac{dF}{ds}\right)^{2} +3F^{3}+6\left(1-\frac{v_{0}}{c_{0}}\right)F^{2}+12C_{1}F+12C_{2}=0,
\label{33}
\end{equation}
where $C_{2}$ is the second integration constant. We choose the integration constant as $C_{2}=0$ and introduce the function $Y(s)$ by relation $F(s)=(\chi h_{0}^{2}/3)Y(s)$ which transforms Eq.(\ref{33}) to the following form, 
\begin{equation} 
\left(\frac{dY}{ds}\right)^{2} +Y^{3}+\frac{6}{\chi h_{0}^{2}}\left(1-\frac{v_{0}}{c_{0}}\right)Y^{2}
+\frac{36C_{1}}{\chi^{2} h_{0}^{4}}Y=0.
\label{34}
\end{equation}
The solution of elliptic differential equation (\ref{34}) yields the function $\zeta(x,t)=(\chi h_{0}^{3}/3)Y(s)$ as
\begin{equation} 
\zeta(x,t)=\Lambda_{0} k^{2}\mathrm{cn}^{2}(W_{0}\xi,k),
\label{35}
\end{equation}
where $\Lambda_{0}$ is an arbitrary positive constant, $\xi=x-x_{0}-v_{0}t$ and $\mathrm{cn}(z,k)$ is the elliptic Jacobi function. The parameters $W_{0}$ and $v_{0}$ in this periodic solution are
\begin{equation} 
W_{0}=\frac{1}{2h_{0}}\sqrt{\frac{3\Lambda_{0}}{\chi h_{0}}},~~~~
v_{0}=c_{0}+\frac{c_{0}\Lambda_{0}}{2h_{0}}(2k^{2}-1).	
\label{36}
\end{equation}
Thus, this periodic solution depends for two positive free parameters as $0< k< 1$ and $\Lambda_{0}>0$. Eqs. (\ref{30}) and (\ref{35}) lead to solution for the function
$\eta(x,t)$ as
\begin{equation} 
\eta(x,t)=\Lambda_{0} k^{2}\mathrm{cn}^{2}(W_{0}\xi,k)
+\frac{\Lambda_{0}^{2}k^{4}}{4h_{0}}\mathrm{cn}^{4}(W_{0}\xi,k).
\label{37}
\end{equation}
This periodic solution differs from known solution of KdV equation by the second term which is not a small for relatively large amplitudes. We introduce here the dimensionless parameter $\delta=\mathrm{max}(|\eta(x,t)|/h_{0})$ which for solution in Eq. (\ref{37}) is
\begin{equation} 
\delta=\frac{\Lambda_{0}}{h_{0}}k^{2}+\frac{\Lambda_{0}^{2}}{4h_{0}^{2}}k^{4}.
\label{38}
\end{equation}
The periodic solution in Eq. (\ref{37}) reduces to the solitary wave for limiting case with $k=1$ as 
\begin{equation} 
\eta(x,t)=\Lambda_{0} \mathrm{sech}^{2}(W_{0}\xi)+\frac{\Lambda_{0}^{2}}{4h_{0}}\mathrm{sech}^{4}(W_{0}\xi),
\label{39}
\end{equation}
where $\xi=x-x_{0}-v_{0}t$ with $v_{0}=c_{0}+c_{0}\Lambda_{0}/2h_{0}$, and the inverse width $W_{0}$ is given in Eq. (\ref{36}).
The periodic solution in Eq. (\ref{37}) for small parameter $k$ ($k\ll 1$) has the form,
\begin{equation} 
\eta(x,t)=\Lambda_{0}k^{2}\cos^{2}(W_{0}\xi)+\frac{\Lambda_{0}^{2}k^{4}}{4h_{0}}\cos^{4}(W_{0}\xi),
\label{40}
\end{equation}
where $W_{0}$ and $v_{0}$ are given in Eq. (\ref{36}).
 
It follows from Eq. (\ref{30}) that the relative difference for soliton solution based on KdV equation [$\eta(x,t)=\Lambda_{0} \mathrm{sech}^{2}(W_{0}\xi)$] and soliton solution for extended KdV equation presented in (\ref{39}) is about $12\%$ for $\delta=0.6$ and $\xi=0$.
The solitary wave given in Eq. (\ref{39}) has for co-moving frame ($x'=x-x_{0}-v_{0}t$ with $v_{0}=c_{0}+c_{0}\Lambda_{0}/2h_{0}$) the following dimensionless form, 
\begin{equation} 
U(S)=A_{0}\mathrm{sech}^{2}(S)+\frac{1}{4}A_{0}^{2}\mathrm{sech}^{4}(S),
\label{41}
\end{equation}
where $U=\eta/h_{0}$, $A_{0}=\Lambda_{0}/h_{0}$ and $S=W_{0}x'$.
\begin{figure}[h]
\includegraphics[width=1\textwidth]{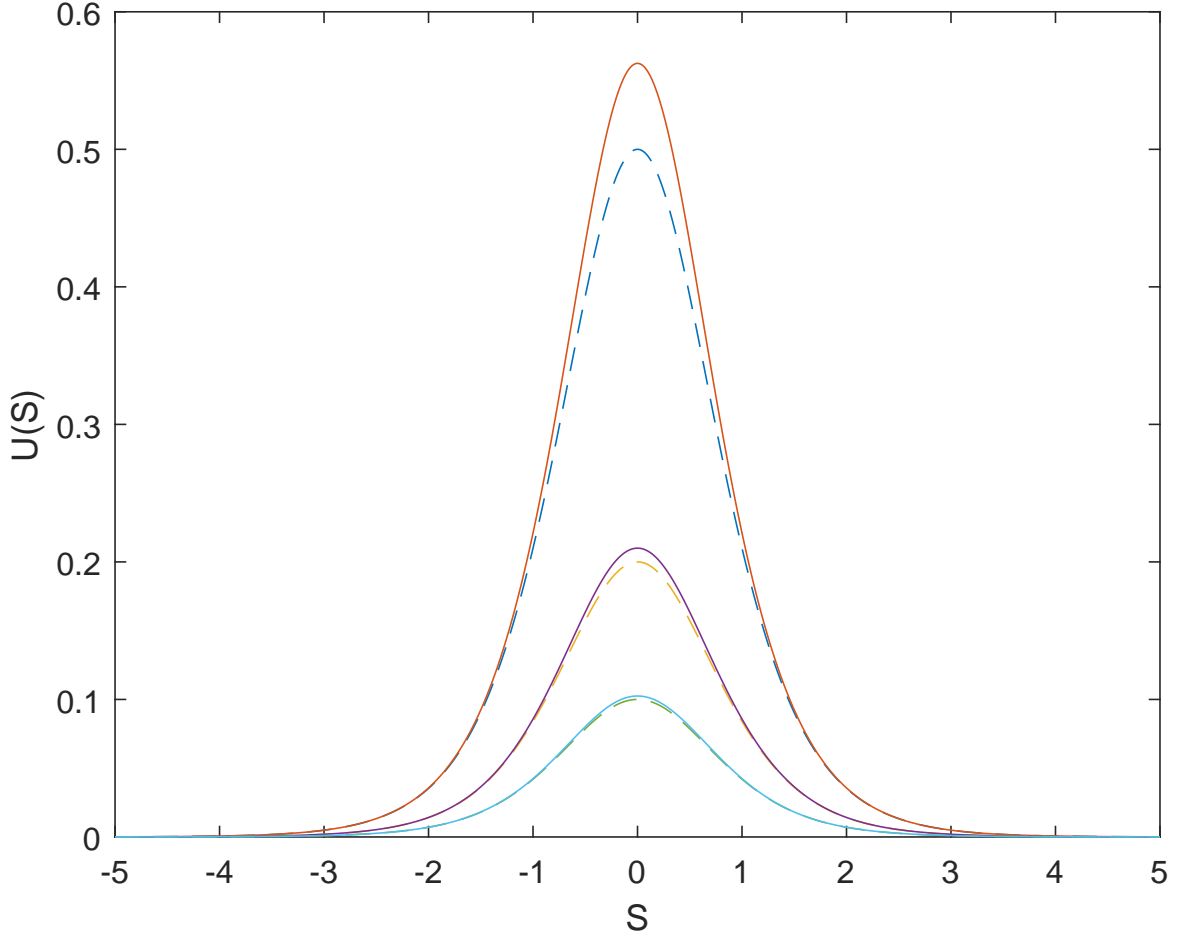}
\caption{The dimensionless profiles of solitary waves given in  Eq. (\ref{41}) are shown by solid lines and the dimensionless profiles  of standard KdV equation (in the co-moving frame) are presented by discontinuous lines. These two profiles are given by pairs for the following parameters: $A_{0}=0.5$, $A_{0}=0.2$, $A_{0}=0.1$. The amplitudes of solitary waves presented by solid and discontinuous lines decrease continuously with decreasing of  parameter $A_{0}$.}
\label{FIG.1.}
\end{figure}
Figure 1 displays the dimensionless profiles for solitary waves (\ref{41}) of extended KdV equation and appropriate solutions  of KdV equation [$U(S)=A_{0}\mathrm{sech}^{2}(S)$] by pairs with solid and discontinuous lines respectively.
These profiles are presented for different values of the amplitude parameter: $A_{0}=0.5$, $A_{0}=0.2$, $A_{0}=0.1$. The amplitudes of solitary waves decreases continuously with decreasing of the parameter $A_{0}$. This figure also demonstrates that the difference for these two solutions increases when the parameter $A_{0}$ grows.

\section{Extended KdV equation for decaying gravity waves}

In this section we generalize the extended KdV equation for describing the decaying effect of propagating waves. Such generalization is connected with additional term $\Gamma \zeta$ in the left side of Eq. (\ref{29}). The explicit form for parameter  $\Gamma$ is derived below using the dimensionless analysis and critical parameters for decaying effect. Thus, the generalized extended KdV equation has the form, 
\begin{equation}
\partial_{t}\zeta+c_{0}\partial_{x}\zeta+\sigma\partial_{x}^{3}\zeta
+\beta\zeta\partial_{x}\zeta+\Gamma\zeta=0,
\label{42}
\end{equation}%
where $\beta=3c_{0}/2h_{0}$,  $\sigma=(\chi/6)c_{0}h_{0}^{2}$, and $\Gamma$ is the parameter describing the decaying effect.

Note that the dimensionless form for Eq. (\ref{42}) follows by introducing new variables $s=(x-c_{0}t)/\sqrt{\chi}h_{0}$ and $\tau=(c_{0}/6\sqrt{\chi}h_{0})t$ and the dimensionless function $\Phi$ as
\begin{equation} 
\Phi(s,\tau)=\frac{9}{h_{0}}\zeta(x,t).	
\label{43}
\end{equation}
In this case the dimensionless extended KdV equation is 
\begin{equation} 
\partial_{\tau}\Phi+\partial_{s}^{3}\Phi+\Phi\partial_{s}\Phi
+\alpha\Phi=0,
\label{44}
\end{equation}
where $\alpha=6\sqrt{\chi}h_{0}\Gamma/c_{0}$. Thus, the function $\eta(x,t)$ given by Eq. (\ref{30}) has the form,
\begin{equation} 
\eta(x,t)=\frac{h_{0}}{9}\left(\Phi(s,\tau)+
\frac{1}{36}\Phi^{2}(s,\tau)\right).		
\label{45}
\end{equation}

We accept that the parameter $\Gamma$ depends on kinematic 
viscosity $\nu$ (momentum diffusivity) and the capillary length $\lambda_{c}$ defined as
\begin{equation} 
\lambda_{c}=\sqrt{\frac{\gamma}{g\rho}}	,
\label{46}
\end{equation}
where $\gamma$ and $\rho$ are the surface tension and mass density respectively. The dimensionless analysis with these two parameters yields 
\begin{equation} 
\Gamma=\frac{Q\nu}{\lambda_{c}^{2}}=\frac{Q\mu g}{\gamma},
\label{47}
\end{equation}
where $Q$ is dimensionless function of temperature and $\mu=\nu\rho$ is viscosity of the fluid. This equation can be confirmed by estimation of the characteristic propagation distances for decaying water waves.
The appropriate parameters for water are $\nu=0.01\mathrm{cm^{2}/s}$ and $\lambda_{c}=0.276\mathrm{cm}$ (for temperature $T=20^{o}\mathrm{C}$) which yields $\Gamma=0.131\times Q~\mathrm{s^{-1}}$.
Using Eq. (\ref{65}) we have the propagation distance for solitons as a function of time: $L(t)=\int_{0}^{t} v(t')dt'$. This yields the propagation distance for solitary waves as
\begin{equation} 
L(t)=c_{0}t+\frac{c_{0}\Lambda_{0}}{2h_{0}\Gamma}\left(1-\exp(-\Gamma t) \right),
\label{48}
\end{equation}
where we assume that the condition $\alpha\ll 1$ is satisfied. Hence, for enough long distances (with $\Gamma t\gg 1$) we have $L(t)=c_{0}t+c_{0}\Lambda_{0}/2h_{0}\Gamma$. The dimensionless function $Q$ can be found by Eqs. (\ref{47}) and (\ref{48}) with the appropriate experimental data for propagating distances of decaying solitary waves given as a function of time.

 \section{Decaying wave solutions for extended KdV equation}

In this section we derive the decaying traveling wave solutions for extended KdV equation (\ref{42}) with the transformation given in Eq. (\ref{30}). Using the techniques of perturbation theory \cite{37} we make the replacement $\Gamma\rightarrow \varepsilon\Gamma$ in Eq. (\ref{42}) where $\varepsilon $ is the dimensionless small parameter $\varepsilon \ll 1$. Thus, we assume here that the condition $\alpha\ll 1$ is satisfied. Note that $\varepsilon $ is the formal parameter which we set in the final stage of calculations as $\varepsilon =1$. The traveling wave for Eq. (\ref{42}) can be written in the form,
\begin{equation} 
\zeta(x,t)=f(\tau)\Psi(\Theta),
\label{49}
\end{equation}
where $\tau=\varepsilon t$ is a slow time, and the variables $\Theta$ is given by
\begin{equation} 
\Theta=G(\tau)X,~~~~X=x-x_{0}-\int_{0}^{t}v(\tau')dt'.
\label{50}
\end{equation}
Here $f(\tau)$, $G(\tau)$ and $v(\tau)$ are some unknown functions of   slow variable $\tau=\varepsilon t$. We assume that $\alpha\ll 1$ and hence the condition $6\sqrt{\chi}h_{0}\Gamma/c_{0}\ll 1$
is satisfied. 
The substitution of Eq. (\ref{49}) to (\ref{42}) yields in zero and first order to parameter $\varepsilon $ the system of equations, 
\begin{equation} 
(c_{0}-v(t))\frac{d\Psi(\Theta)}{d\Theta}+\beta f(t)\Psi(\Theta)\frac{d\Psi(\Theta)}{d\Theta}+
\sigma G^{2}(t)\frac{d^{3}\Psi(\Theta)}{d\Theta^{3}}=0,
\label{51}
\end{equation}
 
\begin{equation} 
\frac{df(t)}{dt}\Psi(\Theta)+f(t)
\frac{dG(t)}{dt}\frac{d\Psi(\Theta)}{d\Theta}
X+\Gamma f(t)\Psi(\Theta)=0,
\label{52}
\end{equation}
where the variable $\Theta$ for $\varepsilon=1$ is
\begin{equation} 
\Theta=G(t)X,~~~~X=x-x_{0}-\int_{0}^{t}v(t')dt'.
\label{53}
\end{equation}

Note that in the last stage of this method we put $\varepsilon=1$, and hence we have $\tau=t$ in Eqs. (\ref{51})-(\ref{53}). 
In Eq. (\ref{51}) the function $\Psi(\Theta)$ and the functions $f(t)$, $v(t)$, $G(t)$ depend on different variables as $\Theta$ and $t$. Thus, it follows from Eq. (\ref{51}) that the necessary conditions for existing of solutions for this equation are
\begin{equation} 
c_{0}-v(t)=af(t),~~~~G^{2}(t)=bf(t),
\label{54}
\end{equation}
where $a$ and $b$ are some constants.
In this case Eqs. (\ref{51}) and (\ref{52}) can be written as
\begin{equation}
\sigma b\frac{d^{3}\Psi}{d\Theta^{3}}+\beta \Psi\frac{d\Psi}{d\Theta}+ 
a\frac{d\Psi}{d\Theta}=0,
\label{55}
\end{equation}
\begin{equation} 
\frac{df(t)}{dt}\Psi(\Theta)+\frac{1}{2}\sqrt{bf(t)}\frac{df(t)}{dt}
\frac{d\Psi(\Theta)}{d\Theta}X+\Gamma f(t)\Psi(\Theta)=0.
\label{56}
\end{equation}
The first and second integration of Eq. (\ref{55}) yields
\begin{equation}
\sigma b\frac{d^{2}\Psi}{d\Theta^{2}} +\frac{\beta}{2}\Psi^{2}+a\Psi+C_{1}=0,
\label{57}
\end{equation}
\begin{equation} 
\sigma b\left(\frac{d\Psi}{d\Theta}\right)^{2} +\frac{\beta}{3}\Psi^{3}+a\Psi^{2}+2C_{1}\Psi+2C_{2}=0,
\label{58}
\end{equation}
where $C_{1}$ and $C_{2}$ are the integration constants.
In the case when $\alpha\ll 1$ we have by Eqs. (\ref{49}), (\ref{56}) and (\ref{58}) (see the Appendix C) the decaying quasi-periodic solution for the functions $\zeta(x,t)$ as
\begin{equation} 
\zeta(x,t)=\Lambda(t)k^{2}\mathrm{cn}^{2}(W(t)X,k).
\label{59}
\end{equation}
In this solution we have the amplitude $\Lambda(t)$, inverse width $W(t)$ and velocity  $v(t)$ as 
\begin{equation} 
\Lambda(t)=\Lambda_{0}\exp(-\Gamma t), 
\label{60}
\end{equation}
\begin{equation} 
W(t)=\frac{1}{2h_{0}}\sqrt{\frac{3\Lambda(t)}{\chi h_{0}}},~~~~
v(t)=c_{0}+\frac{c_{0}\Lambda(t)}{2h_{0}}(2k^{2}-1).	
\label{61}
\end{equation}
The variable $X=x-x_{0}-\int_{0}^{t}v(t')dt'$ has an explicit form,
\begin{equation} 
X=x-x_{0}-c_{0}t-\frac{c_{0}\Lambda_{0}}{2h_{0}\Gamma}(2k^{2}-1)[1-
\exp(-\Gamma t)].
\label{62}
\end{equation}
Eqs. (\ref{30}) and (\ref{59}) lead to the function $\eta(x,t)$ as
\begin{equation} 
\eta(x,t)=\Lambda(t)k^{2}\mathrm{cn}^{2}(W(t)X,k) + \frac{\Lambda^{2}(t)k^{4}}{4h_{0}}
\mathrm{cn}^{4}(W(t)X,k).
\label{63}
\end{equation}
The decaying soliton solution follows from Eq. (\ref{63}) with parameter $k=1$ as
\begin{equation} 
\eta(x,t)=\Lambda(t)\mathrm{sech}^{2}(W(t)X)+\frac{\Lambda^{2}(t)}{4h_{0}}\mathrm{sech}^{4}(W(t)X),
\label{64}
\end{equation}
where the functions $v(t)$ and $X$ are
\begin{equation} 
v(t)=c_{0}+\frac{c_{0}\Lambda(t)}{2h_{0}},~~~~X=x-x_{0}
-c_{0}t-\frac{c_{0}\Lambda_{0}}{2h_{0}\Gamma}[1-\exp(-\Gamma t)].	
\label{65}
\end{equation}
In the case with $k^{2}\ll 1$ we have by Eq. (\ref{63}) the periodic solution as
\begin{equation} 
\eta(x,t)=\Lambda(t)k^{2}\cos^{2}(W(t)X)+\frac{\Lambda^{2}(t)k^{4}}{4h_{0}}\cos^{4}(W(t)X).
\label{66}
\end{equation}
We emphasis that in the limit $\Gamma\rightarrow 0$ the solutions in Eqs. (\ref{63}), (\ref{64}) and (\ref{66}) coincide with appropriate solutions in Eqs. (\ref{37}), (\ref{39}) and (\ref{40}).   

\section{Discussion}

There is a simple and important connection between traveling solutions of extended KdV  Eq. (\ref{29}) and (\ref{42}). Let us apply the transformation $\Lambda_{0}\mapsto \Lambda(t)$ to traveling solutions defined in Eqs. (\ref{35})-(\ref{40}). In this case the parameters $W_{0}$ and $v_{0}$ given in Eq. (\ref{36}) yield the functions $W(t)$ and $v(t)$ defined by Eq. (\ref{61}). Thus, the transformation $\Lambda_{0}\mapsto \Lambda(t)$ leads the following mapping:
\begin{equation} 
W_{0}\mapsto W(t),~~~~v_{0}\mapsto v(t).
\label{67}
\end{equation}
\begin{figure}[h]
\includegraphics[width=1\textwidth]{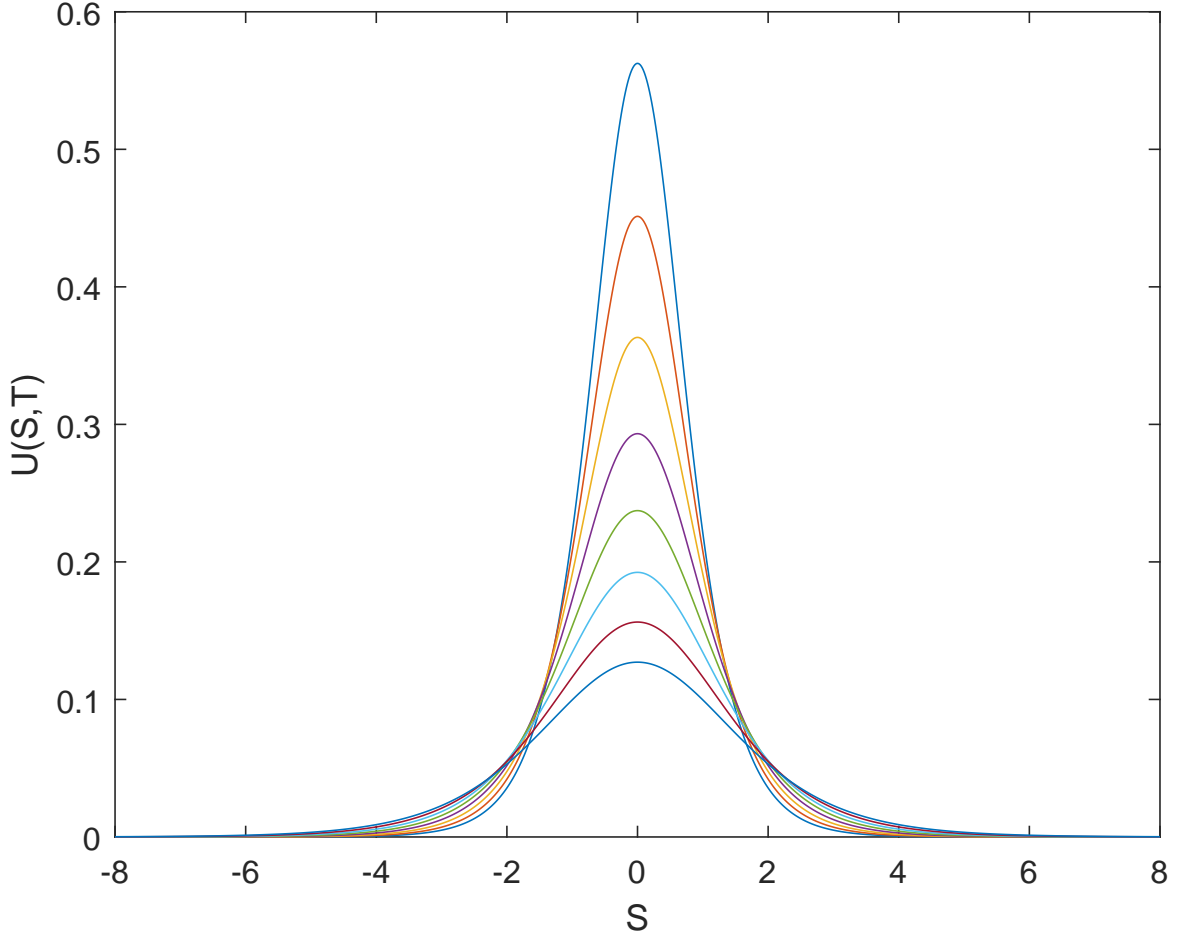}
\caption{The dimensionless profiles $U(S,T)$ of solitary waves (in co moving frame) given in Eq. (\ref{70}) for parameter $A_{0}=0.5$ and fixed dimensionless times $T=0.2n$ where $n=0, 1,...,7$. The amplitudes of solitary waves decrease and the width increase continuously with increasing of the dimensionless time $T$.}
\label{FIG.2.}
\end{figure}
Note that the variable $\xi$ used in Eq. (\ref{35}) can be written as
\begin{equation} 
\xi=x-x_{0}-\int_{0}^{t}v_{0}dt'.	
\label{68}
\end{equation}
Hence, the mapping $v_{0}\mapsto v(t)$
in Eq. (\ref{68}) yields the transformation $\xi\mapsto X$ where the function $X$ is defined by Eq. (\ref{53}). Thus, we have shown that the transformation $\Lambda_{0}\mapsto \Lambda(t)$ applied to traveling solutions in Eqs. (\ref{35}), (\ref{37}), (\ref{39}) and (\ref{40}) leads the decaying traveling solutions in Eqs. (\ref{59}), (\ref{63}), (\ref{64}) and (\ref{66}) respectively. We emphasize that the found connection  between traveling solutions of extended KdV Eq. (\ref{29}) and (\ref{42}) occur when the condition $\alpha\ll 1$ is satisfied.

We introduce the co-moving frame for solitary wave as
\begin{equation} 
x'=x-x_{0}-c_{0}t-\frac{c_{0}\Lambda_{0}}{2h_{0}\Gamma}[1-\exp(-\Gamma t)].
\label{69}
\end{equation}
In this case Eq. (\ref{64}) for solitary wave in co-moving frame has the dimensionless form,   
\begin{equation} 
U(S,T)=A_{0}e^{-T}\mathrm{sech}^{2}(Se^{-T/2})+\frac{1}{4}A_{0}^{2}e^{-2T}\mathrm{sech}^{4}
(Se^{-T/2}),
\label{70}
\end{equation}
where $U=\eta/h_{0}$, $A_{0}=\Lambda_{0}/h_{0}$, $T=\Gamma t$, and $S=W_{0}x'$ ($W_{0}=W(0)$).

The profiles of dimensionless solitary waves given in Eq. (\ref{70}) are shown in Fig. 2 for parameter $A_{0}=0.5$ and fixed dimensionless times $T=0.2n$ with $n=0, 1,...,7$.
The amplitudes of these solitary waves decrease and the width increase continuously with increasing of the dimensionless time $T$ or the number $n$.

\section{Conclusion}

In this paper, we have derived the extended KdV equation for the water waves with arbitrary amplitudes. The only restriction to the surface deviation is connected with the stability condition for the waves. It is used in this derivation of extended KdV equation the long-wave approximation given by the condition as $\epsilon^{2}\ll 1$. Moreover, we have generalized the extended KdV equation adding the term describing the decaying effect of the waves. The decaying effect is important for describing the propagation of the waves to long distances. It is shown that the term describing decaying effect in the extended KdV equation depends on two parameters as the kinematic viscosity $\nu$ (momentum diffusivity) and the capillary length $\lambda_{c}$. The explicit form for the decaying term is derived using the dimensionless analysis with the critical parameters $\nu$ and $\lambda_{c}$. Hammack and Segur have demonstrated in their paper \cite{32} that the agreement to within about 20$\%$ is observed over the entire range of experiments examined for moderate amplitudes of the waves. It is remarkable that the difference of solutions for the extended and standard KdV equatins with enough large stable amplitudes is also within the same range about 20$\%$. Hence, we hope that the approach based on the extended KdV equation can significantly improve the accuracy of theory for long gravity waves in incompressible fluid. Thus, we conclude that the additional and more detail comparison of new theory with experimental data for gravity waves is important field for future studies.      

We have also found a set of periodic, quasi-periodic and solitary wave solutions for extended KdV equation in the cases of non-decaying and decaying waves. We have demonstrated in the Appendix B that the fundamental approaches based on the inverse scattering method are applicable for solving the extended KdV equation in the cases when the decaying effect is negligibly small. Such cases always occur for restricted propagation distances of the waves. Thus, in these cases the solutions of extended KdV equation can be found by inverse scattering method or Gel'fand-Levitan-Marchenko integral equation. In conclusion, we have derived the extended KdV equation for gravity waves which is generalizing the theory based on the KdV equation. This new approach to long gravity waves has no strong restrictions on the wave's amplitude. As we have mentioned, the only limitation to wave amplitudes is connected with stability condition for solutions of the extended KdV equation.

\appendix

\section{Euler equations with long wave approximation}

We have defined the following dimensionless variables:
$\tau=c_{0}t/l$, $\lambda=x/l$ and $\xi=z/h_{0}$. In this case
the dimensionless conservation Eq. (\ref{3}) can be written as
\begin{equation}
\partial_{\lambda}\tilde{u}+\partial_{\xi}\tilde{w}=0,
\label{1a}
\end{equation}
where $u(x,t)=c_{0}\tilde{u}(\lambda,\tau)$ and $\tilde{u}$ is the dimensionless velocity. It follows from Eq. (\ref{1a}) that the dimensionless velocity $\tilde{w}$ is given by relation $w(x,z,t)=\epsilon c_{0}\tilde{w}(\lambda,\xi,\tau)$ where $\epsilon=h_{0}/l$. We use these dimensionless variables and functions for derivation of the general equation for pressure $P$. We show below that full pressure is the sum of static and dynamic pressures.
\newline

\textit{1. Dynamic pressure}

The Euler Eq. (\ref{2}) can be written in the standard form as
\begin{equation}
\frac{Dw}{Dt}=-\frac{1}{\rho}\partial_{z}P-g.
\label{2a}
\end{equation}%
Using defined dimensionless variables and functions we can write this equation in the form,
\begin{equation}
\epsilon^{2}\frac{D\tilde{w}}{D\tau}=-\frac{1}{g}\left(\frac{1}{\rho}\partial_{z}P+g \right).
\label{3a}
\end{equation}%
Hence, for long waves approximation $\epsilon^{2}\ll 1$ we have the following equation for pressure $P$:
\begin{equation}
\frac{1}{\rho}\partial_{z}P+g=0.
\label{4a}
\end{equation}
We present the full pressure $P$ as the sum of two terms,
\begin{equation} 
P=P_{g}+P_{d},
\label{5a}
\end{equation}
where $P_{g}$ and $P_{d}$ are the static and dynamic gravitational pressures. We note that dynamic gravitational pressures $P_{d}$ is necessary for correct description of the dispersion relation in the first order to small parameter $\epsilon^{2}$. The static gravitation pressure $P_{g}$ depends on the liquid depth $h(x,t)$ and the vertical coordinate $z$, and the dynamic gravitational pressure $P_{d}$ depends on variables $x$ and $t$ only which follows from the linear differential equation (\ref{14}). Eqs. (\ref{4a}) and (\ref{5a}) yield the equation for static pressure as 
\begin{equation}
\frac{1}{\rho}\partial_{z}P_{g}=-g,
\label{6a}
\end{equation}
because $\partial_{z}P_{d}=0$.
This equation leads to the following static gravitation pressure: 
\begin{equation} 
P_{g}=P_{0}+\rho g[h(x,t)-z],	
\label{7a}
\end{equation}
where $z$ is the vertical coordinate and $P_{0}$ is the pressure at $z=h$. 
It is shown in Sec. II that the dynamic pressure $P_{d}$ is given by Eq. (\ref{17}) as
\begin{equation} 
P_{d}=\rho\sigma\partial_{x}^{2}u=\frac{\chi}{6}\rho gh_{0}^{2}\partial_{x}^{2}\zeta,
\label{8a}
\end{equation}
where $\zeta(x,t)=(c_{0}/g)u(x,t)$.

We note that the dynamic pressures $P_{d}$ is also necessary for implementation of the function $\mathcal{D}(x,t)\equiv\rho^{-1}\partial_{x}P_{d}(x,t)$ given in explicit form by Eq. (\ref{17}). The Euler Eq. (\ref{1}) with the defined above function $\mathcal{D}(x,t)$ yields Eq. (\ref{6}) which is the main equation for developed here theory. Derivation of the extended KdV equation for long gravity waves is also based on the transformation given in Eq. (\ref{8}). The detail description and derivation of the extended KdV equation is presented in Sec. II. 
\newline

\textit{2. Conservation equation}

Integration of the conservation Eq. (\ref{3}) yields
\begin{equation}
\int_{0}^{h}(\partial_{x}u+\partial_{z}w)dz=0.
\label{9a}
\end{equation}%
We have the apparent boundary conditions as $[w]_{z=0}=0$ and $[w]_{z=h}=0$. Thus, Eq. (\ref{9a}) can be written as
\begin{equation}
\partial_{x}\int_{0}^{h}udz-[u]_{z=h}\partial_{x}h=0.
\label{10a}
\end{equation}%
Considering the boundary condition $Dh/Dt=\partial_{t}h+[u]_{z=h}\partial_{x}h=0$ and the velocity $u=u(x,t)$ which does not depend on variable $z$ we have by Eq. (\ref{10a}) the following conservation equation,
\begin{equation}
\partial_{t}h+\partial_{x}(uh)=0.
\label{11a}
\end{equation}%
This is well-known conservation equation for the gravity waves in shallow water. 
\newline

\textit{3. Riemann invariant}

We emphasize that the term $\mathcal{D}(x,t)$ in the system of Eqs. (\ref{6}) and (\ref{7})
is proportional to $\epsilon^{2}$ which follows from Eq. (\ref{19}). Thus, in the limit when
$\epsilon^{2}$ tends to zero we have Eqs. (\ref{6}) and (\ref{7}) with $\mathcal{D}(x,t)\equiv 0$.
The system of Eqs. (\ref{6}) and (\ref{7}) with $\mathcal{D}(x,t)\equiv 0$ has the Riemann invariant as
\begin{equation}
u(x,t)=2\sqrt{gh_{0}+g\eta(x,t)}-2\sqrt{gh_{0}}.	
\label{12a}
\end{equation}
This Riemann invariant transforms Eqs. (\ref{6}) and (\ref{7}) with $\mathcal{D}(x,t)\equiv 0$ to a single equation,
\begin{equation}
\partial_{t}u+c_{0}\partial_{x}u
+\frac{3}{2}u\partial_{x}u=0.
\label{13a}
\end{equation}

Using the method of characteristics we can present the general solution of Eq. (\ref{13a}). The general solution $u(x,t)$ of Eq. (\ref{13a}) with initial condition $u(x,0)=U_{0}(x)$ is
\begin{equation}
u(x,t)=U_{0}(\xi),~~~~x=c_{0}t+\frac{3}{2}U_{0}(\xi)t+\xi.
\label{14a}
\end{equation}
Here $\xi$ is the parameter of this parametric solution which can be excluded from the algebraic system of equations given in Eq. (\ref{14a}). Let us the initial condition for the surface deviation is given as $\eta(x,0)=F_{0}(x)$ then the function $U_{0}(x)$ is
\begin{equation}
U_{0}(x)=2\sqrt{gh_{0}+gF_{0}(x)}-2\sqrt{gh_{0}}.	
\label{15a}
\end{equation}
The solution for surface deviation $\eta(x,t)$ is given by Eq. (\ref{12a}) as $\eta(x,t)=(c_{0}/g)u(x,t)+(1/4g)u^{2}(x,t)$ where the velocity $u(x,t)$ is defined by Eq. (\ref{14a}).

The solution of the system of Eqs. (\ref{6}) and (\ref{7}) with 
$\mathcal{D}(x,t)\equiv 0$ can also be presented in another form. The second equation in (\ref{14a}) yields $\xi=f(x,t)$ where $f(x,t)$ is some function of variables $x$ and $t$, then we have the velocity as $u(x,t)=U_{0}(f(x,t))$.
Hence, the surface deviation $\eta(x,t)$ is given by 
\begin{equation}
\eta(x,t)=\frac{c_{0}}{g}U_{0}(f(x,t))+\frac{1}{4g}U_{0}^{2}(f(x,t)).
\label{16a}
\end{equation}
We note that in general case the function $\xi=f(x,t)$ has not a single value of $\xi$ for all values of time $t$ because the projections of two characteristics in Eq. (\ref{14a}) to the plane $(x,t)$ can cross for some values of time $t$. It follows from Eq. (\ref{16a}) that in general case the function $\eta(x,t)$ has not a single value for all values of time $t$ as well.
\newline

\section{Exact solutions for extended KdV equation}

The dimensionless form for Eq. (\ref{29}) follows by introducing new variables $s=(x-c_{0}t)/\sqrt{\chi}h_{0}$ and $\tau=(c_{0}/6\sqrt{\chi} h_{0})t$ and the dimensionless function as 
\begin{equation} 
\mathcal{Z}(s,\tau)=\frac{3}{2h_{0}}\zeta(x,t)=
\frac{3}{2c_{0}}u(x,t).	
\label{1b}
\end{equation}
In this case the dimensionless extended KdV equation is 
\begin{equation} 
\partial_{\tau}\mathcal{Z}+\partial_{s}^{3}\mathcal{Z}+6\mathcal{Z}\partial_{s}\mathcal{Z}=0.
\label{2b}
\end{equation}
The function $\eta(x,t)$ is given by Eq. (\ref{30}) as
\begin{equation} 
\eta(x,t)=\frac{2h_{0}}{3}\mathcal{Z}(s,\tau)+\frac{h_{0}}{9}\mathcal{Z}^{2}(s,\tau).	
\label{3b}
\end{equation}
Eq. (\ref{2b}) is Galilean invariant, i.e., it is unchanged by the transformation $\tilde{\mathcal{Z}}(s',\tau')=\mathcal{Z}(s,\tau)-c/6$ where $\tau'=\tau$ and $s'=s-c\tau$.

One-soliton solution of Eq. (\ref{2b}) is
\begin{equation} 
\mathcal{Z}(s,\tau)=A\mathrm{sech}^{2}[\sqrt{A/2}(s-2A\tau-s_{0})],	
\label{4b}
\end{equation}
where $A$ and $s_{0}$ are an arbitrary real constants. 
Two-soliton solution has the form,
\begin{equation} 
\mathcal{Z}(s,\tau)=2\partial_{s}^{2}\ln\left[1+b_{1}\exp(\phi_{1})+b_{2}
\exp(\phi_{2})+b_{0}b_{1}b_{2}\exp(\phi_{1}+\phi_{2})\right], 	
\label{5b}
\end{equation}
where $\phi_{1}=a_{1}s-a_{1}^{3}\tau$, $\phi_{2}=a_{2}s-a_{2}^{3}\tau$
and $b_{0}=(a_{1}-a_{2})^{2}/(a_{1}+a_{2})^{2}$.

Eq. (\ref{2b}) has also the algebraic soliton solutions as
\begin{equation} 
\mathcal{Z}(s,\tau)=-\frac{6s(s^{3}-24\tau)}{(s^{3}+12\tau)^{2}},	
\label{6b}
\end{equation}
\begin{equation} 
\mathcal{Z}(s,\tau)=2\partial_{s}^{2}\ln\left(s^{6}+60s^{3}\tau
-720\tau^{2}\right). 	
\label{7b}
\end{equation}

The inverse scattering method for Eq. (\ref{2b}) leads to solutions in the form, 
\begin{equation} 
\mathcal{Z}(s,\tau)=2\partial_{s}^{2}K(s,s;\tau), 	
\label{8b}
\end{equation}
where the function $K(s,q;\tau)$ is a solution of Gel'fand-Levitan-Marchenko integral equation:
\begin{equation} 
K(s,q;\tau)+F(s,q;\tau)+\int_{s}^{\infty}K(s,p;\tau)F(p,q;\tau)dp=0.	
\label{9b}
\end{equation}
The time $\tau$ in this equation is an arbitrary parameter.
Here $F(s,q;\tau)$ is an arbitrary function which rapidly decreases
for $s\rightarrow +\infty$ and satisfying the linear equations:
\begin{equation} 
\partial_{s}^{2}F-\partial_{q}^{2}F=0,~~~~\partial_{\tau}F
+(\partial_{s}+\partial_{q})^{3}F=0. 	
\label{10b}
\end{equation}
Thus, every function $F(s,q;\tau)$ satisfying these equations and appropriate decreasing condition generates a solution of Eq. (\ref{2b}) by Eqs. (\ref{8b}) and (\ref{9b}).

\section{Decaying waves}

In this Appendix we consider the decaying quasi-periodic solution of Eqs. (\ref{56}) and (\ref{58}). We take in Eq. (\ref{58}) the second integration constant as $C_{2}=0$ which yields the solution,
\begin{equation} 
\Psi({\Theta})=\Lambda_{0}k^{2}\mathrm{cn}^{2}(BG(t)X,k),
\label{1c}
\end{equation}
where the parameters $\Lambda_{0}$, $B$ and function $G(t)$ are
\begin{equation} 
\Lambda_{0}=\frac{6\sigma D_{0}}{\beta}=\frac{2}{3}\chi h_{0}^{3}D_{0},
~~~~D_{0}\equiv \frac{1}{\sigma}\sqrt{\frac{a^{2}}{4}-\frac{2\beta C_{1}}{3}},	
\label{2c}
\end{equation}
\begin{equation} 
B=\sqrt{D_{0}/2b},~~~~G(t)=\sqrt{bf(t)}.
\label{3c}
\end{equation}
In this solution the modulus of elliptic Jacobi function $k$ is connected with parameter $a$ as
\begin{equation} 
a =-2\sigma D_{0}(2k^{2}-1)=-\frac{c_{0}\Lambda_{0}}{2h_{0}}(2k^{2}-1),
\label{4c}
\end{equation}
with $0< k< 1$.
We define the function $W(t)$ for the wave solution in Eq. (\ref{1c}) as
\begin{equation} 
W(t)\equiv BG(t)=\sqrt{\frac{D_{0}f(t)}{2}},
\label{5c}
\end{equation}
where $D_{0}=3\Lambda_{0}/2\chi h_{0}^{3}$.
Thus, we have shown that the solution in Eq. (\ref{1c}) does not depend on parameter $b$. Using Eqs. (\ref{49}) and (\ref{1c}) we can write the solution of Eq. (\ref{42}) as 
\begin{equation} 
\zeta(x,t)=\Lambda_{0}f(t)k^{2}\mathrm{cn}^{2}(W(t)X,k),
\label{6c}
\end{equation}
where $\Lambda_{0}$ is an arbitrary parameter and the functions $W(t)$, $X$ and $v(t)$ are
\begin{equation} 
W(t)=\frac{1}{2h_{0}}\sqrt{\frac{3\Lambda_{0}f(t)}{\chi h_{0}}},~~~~
~~~~X=x-x_{0}-\int_{0}^{t}v(t')dt',
\label{7c}
\end{equation}
\begin{equation} 
v(t)=c_{0}-af(t)=c_{0}+\frac{c_{0}\Lambda_{0}f(t)}{2h_{0}}(2k^{2}-1).\label{8c}
\end{equation}
Eq. (\ref{56}) has nontrivial  solution only in the case when second term in this equation is zero. Thus, we can consider Eq. (\ref{56}) in the limit $b\rightarrow +0$ which leads to solution as $f(t)=f_{0}\exp(-\Gamma t)$. It is important that this limit does not change the solution presented in Eq. (\ref{6c}) because the functions given in Eqs. (\ref{7c}) and (\ref{8c}) are not depended on parameter $b$.
We note that in Eqs. (\ref{6c}), (\ref{7c}) and (\ref{8c}) the function $f(t)$ is multiplied to an arbitrary parameter $\Lambda_{0}$. Hence, without loss of generality we can take $f_{0}=1$. In this case the functions $f(t)$ and $X$ are 
\begin{equation} 
f(t)=\exp(-\Gamma t),	
\label{9c}
\end{equation}
\begin{equation} 
X=x-x_{0}-c_{0}t-\frac{c_{0}\Lambda_{0}}{2h_{0}\Gamma}(2k^{2}-1)[1-
\exp(-\Gamma t)].
\label{10c}
\end{equation}

\end{document}